\documentclass{aa}
\usepackage{txfonts}
\usepackage{graphicx}

\newcommand{\kms}{km s$^{-1}$}
\newcommand{\flux}{erg s$^{-1}$ cm$^{-2}$}

\newcommand{\clsA}{XLSSC~001} 
\newcommand{\clsB}{XLSSC~002} 
\newcommand{\clsC}{XLSSC~003} 
\newcommand{\clsD}{XLSSC~004} 
\newcommand{\clsE}{XLSSC~005} 

\begin{document}

  \title{The XMM-LSS Survey} \subtitle{First high redshift galaxy
  clusters: relaxed and collapsing systems\thanks{Based on
  observations obtained with XMM-Newton, CFHT, ESO (program ID:
  70.A-0283)}}

\author{I.~Valtchanov\inst{1,}\inst{2} \and
  M.~Pierre\inst{1} \and
  J.~Willis\inst{3}\thanks{Present address: Dept. of Physics and Astronomy, Univ. of Victoria, Victoria, Canada} \and
  S.~Dos Santos\inst{1} \and
  L.~Jones\inst{4} \and
  S.~Andreon\inst{5} \and
  C.~Adami\inst{6} \and
  B.~Altieri\inst{7} \and
  M.~Bolzonella\inst{8} \and
  M.~Bremer\inst{9} \and
  P.-A.~Duc\inst{1} \and
  E.~Gosset\inst{10} \and
  C.~Jean\inst{10} \and
  J.~Surdej\inst{10}}

\institute{
  CEA/Saclay, Service d'Astrophysique, F-91191, Gif-sur-Yvette, France \and
  Astrophysics Group, Blackett Laboratory, Imperial College, London SW7 2BW, UK \and
  ESO, Ave. Alonso de Cordova 3107, Casilla 19, Santiago
  19001, Chile \and
  School of Physics and Astronomy, University of Birmingham, Birmingham, UK \and
  INAF-Osservatorio Astronomico di Brera, via Brera 28, 20121 Milano, Italy  \and
  LAM, Traverse du Siphon, 13012 Marseille, France \and
  ESA, Villafranca del Castillo, Spain \and
  Istituto di Astrofisica Spaziale e Fisica Cosmica, Sezione di
  Milano, via Bassini 15, 20133 Milano \and
  Department of Physics, University of Bristol, H.H. Wills Laboratory, Bristol, UK \and
  Universit\'e de Li\`ege, All\'ee du 6 Ao\^ut, 17, B5C, 4000 Sart Tilman, Belgium}

\offprints{Ivan Valtchanov, \\ \email{i.valtchanov@imperial.ac.uk}}

\date{Received, \today, Accepted, \today}

\titlerunning{XMM-LSS. High redshift clusters.}
\authorrunning{Valtchanov et al.}

\abstract{ We present five newly found galaxy clusters at $z>0.6$ from
  the XMM Large-Scale Structure Survey (XMM-LSS).  All five objects
  are extended X-ray sources in the XMM images. For three of them we
  have sufficient spectroscopically confirmed member galaxies that an
  estimate of the velocity dispersion is possible: XLSSC~001 at
  $z=0.613$ and $\sigma_V=867^{+80}_{-60}$ \kms, XLSSC~002 at
  $z=0.772$ and $\sigma_V=524^{+267}_{-116}$ \kms and XLSSC~003 at
  $z=0.839$ and $\sigma_V=780^{+137}_{-75}$ \kms. These three clusters
  have X-ray bolometric luminosities $L_X \sim 1-3 \times 10^{44}$ erg
  s$^{-1}$ and temperatures $2-4$ keV, and consequently are less
  massive than previously known clusters at similar redshifts, but
  nevertheless they follow the low redshift scaling relations between
  $L_X,\ T$ and $\sigma_v$, within the limits of the measurement
  errors. One of the clusters, XLSSC~004, is detected independently as
  an overdensity of galaxies of a colour R-z'=1.4 that matches the
  redshift of the central galaxy $z=0.87$, although it cannot
  unambiguously be confirmed by the spectroscopic observations alone.
  The highest redshift candidate cluster pertaining to this paper,
  XLSSC~005, is most likely a double cluster complex at a redshift
  around unity, associated with an extended X-ray source with probable
  substructure.

   }

\maketitle

\section{Introduction}

Statistical samples of distant galaxy clusters can serve as a powerful
tool to study the large-scale structure of the Universe. To obtain
cluster samples deep enough and over a large area with optical or
near-infrared observations is quite a difficult task (e.g. Gunn et
al. \cite{gun86}, Postman et al. \cite{pos96}, Gonzales et
al. \cite{gon01}, Stanford et al.\cite{sta97}, Dickinson
\cite{dic97}). Projection effects and the small field-of-view of the
current NIR imagers interplay to make relatively large contiguous
surveys of distant clusters a dubious game.

On the other hand, a much more efficient strategy is to perform
cluster searches in X-rays: galaxy clusters are one of the most
powerful X-ray emitting objects, their X-ray emission is of thermal
origin (free-free emission of the electrons from the gas trapped into
the deep gravitational potential well of the cluster and, in general,
in hydrostatic equilibrium) and depends on the square of the gas
density and so projection effects are not very important. Another
advantage is the fact that high galactic latitude fields of the X-ray
sky are populated mostly by two distinct kinds of object: AGNs or
QSOs, which appear as point-like, and galaxy clusters, which are
extended. It is straightforward to distinguish between the two
  classes using X-ray telescopes with good spatial and spectral
  resolution. Namely, galaxy clusters are extended sources radiating
  by thermal bremsstrahlung with spectra declining steeply with energy
  and so (depending on the temperature) the major part of the cluster
  emission is in the soft X-ray band (below 2 keV). On the other hand,
  AGNs/QSOs are point-like sources and generally have harder power-law
  spectra.  The spectral distinction, however requires sufficient
photon statistics and is prone to many assumptions and uncertainties.

Ever since the first imaging X-ray observations, galaxy clusters were
one of the main targets. This has led to flourishing X-ray based
cluster surveys, initiated by the Einstein Medium Sensitivity Survey
(Gioia et al. \cite{emss}) and then continued with ROSAT both at low
redshifts (e.g. Ebeling et al. \cite{ebe98}, \cite{ebe00}; De Grandi
et al. \cite{deg99}; B\"ohringer et al. \cite{reflex}) and high
redshifts (e.g. Rosati et al 1998, Jones et al. \cite{jon98}, Romer et
al. \cite{rom00}, Ebeling et al. \cite{ebe01}, Gioia et
al. \cite{gio01}, Burke et al. \cite{bur03}, Mullis et
al. \cite{mul03}; for a review see Rosati et al. \cite{ros02}). From
these cluster surveys we have gained knowledge about individual
objects, scaling relations, studies of the physics of the
intra-cluster medium, the large-scale distribution, the clustering
properties and the evolution of galaxy clusters.

Nowadays cluster searches and surveys are brought to a new level with
Chandra and XMM. Some of these projects are a natural continuation of
existing surveys from the previous X-ray missions, following-up
interesting objects and to get better insights on cluster physics,
while other surveys use the publicly available archives of XMM and
Chandra X-ray observations -- this is really an inexpensive way to get
statistically significant number of serendipitously detected clusters
at all redshifts (e.g. Romer et al. \cite{rom01}, Schwope et
al. \cite{sch03}, Kim et al. \cite{kim03}).

The increased sensitivity of XMM and Chandra X-ray observatories,
however, can also be exploited to survey large contiguous areas of the
sky with relatively short exposure times. A fainter cluster/group
population at intermediate redshifts that was not accessible to the
previous X-ray missions will thus be detected.  In return, this will
give us a unique view of the cosmic web at the mass scale traced by
this population.

This is the approach of the XMM Large-Scale Structure Survey (XMM-LSS,
Pierre et al. \cite{msngr}, \cite{estec}). The final objective
is to map a contiguous region of the sky and to study the large-scale
distribution and the clustering properties of the matter traced by the
galaxy clusters and the QSO/AGN population. XMM-LSS is based on XMM
observations and a subsequent multi-wavelength follow-up
programme. The target XMM-LSS survey geometry and depth were chosen
such that to have a statistically significant number of clusters so
that the two-point correlation function of clusters at $0<z<0.5$ and
$0.5<z<1$ can be calculated with a precision reaching that of the
ROSAT-ESO Flux Limited X-ray Survey (REFLEX, Collins et
al. \cite{col00}) which covers a much larger area but only goes out to
$z<0.2$.

To achieve this goal we need a good X-ray cluster detection
procedure. Using extensive XMM image simulations we have developed an
efficient pipeline procedure that uses the full XMM instrument
sensitivity and indeed allows us to detect extended sources down to
flux levels of $5\times 10^{-15}$ \flux\ for 10 ks exposures
(Valtchanov et al. \cite{val01}, Refregier et al. \cite{ref02}). The
successful detection is quite crucial because even the mere counting
of clusters at different redshifts can be used to constrain
cosmological parameters (e.g. Mathiesen \& Evrard \cite{mat98}, Henry
\cite{hen00}, Borgani et al. \cite{bor01}, Refregier et al. \cite{ref02}). 
These cosmoslogical constraints that come from cluster abundance
evolution are complementary and independent to those from cosmic
microwave background and supernovae studies.

Each detected candidate cluster from the pipeline, depending on its
estimated redshift by photometric redshifts, is programmed to a
spectroscopic follow-up ($z<1$) or NIR observations ($z>1$). The
spectroscopic follow-up of the first sample of candidate clusters in
the XMM-LSS at $z<1$ was programmed for observations on Las
Campanas/Magellan and on the ESO/VLT telescopes. The subject of this
paper is to present the first results for a sample of 5 clusters at
$z>0.6$ while the low redshift sample is presented elsewhere (Willis
et al., in preparation).

The paper is organised as follows: first we present the X-ray data
reduction and source detection (Sect.~\ref{sec:x}), then we describe
in Sect.~\ref{sec:opt} the optical identification, spectroscopic
target selection procedure and the observations. In
Sect.~\ref{sec:results}, we present the data analysis results from the
spectroscopic and X-ray observations. Next we discuss each individual
object (Sect.~\ref{sec:obj}) and we end up with the conclusions
(Sect.~\ref{sec:end}). Except where is mentioned, we use $\Lambda$CDM
cosmology ($H_0=70$ km s$^{-1}$ Mpc$^{-1}$, $\Omega_m=0.3,\
\Omega_{\Lambda}=0.7$) for all cosmologically dependent
parameters. All X-ray luminosities are bolometric.

\section{X-ray data}
\label{sec:x}

The candidate clusters for the first spectroscopic run were chosen
from all XMM-LSS pointings received by August 2002. This includes 15
AO-1 pointings of 10 ks exposure and another 15 exposures of 20 ks
from the Guaranteed Time XMM Medium Deep Survey (XMDS). All
observations were of good quality, except two fields that suffered
from high background contamination affecting more than 50\% of the
exposure time.

\subsection{Data reduction}

A detailed description of the pipeline used in the XMM-LSS data
reduction will be presented elsewhere (Pacaud et al., in preparation).
Here we just briefly mention the main steps. The raw X-ray
observations (ODFs) are reduced by the standard XMM Science Analysis
System (XMM-SAS) tasks {\tt emchain} and {\tt epchain} for MOS and PN
detectors respectively. High background periods, related to soft
protons, are excluded from the event lists and raw photon images with
a given pixel scale ($2.5\arcsec$/pixel in this case) in different
energy bands are then created. Subsequently the raw images for each
instrument are filtered using ``\`a trous'' (with holes) iterative
wavelet technique with a Poisson noise model and a threshold of
$10^{-4}$ (equivalent to $3.7\sigma$ in the Gaussian case) for the
significant wavelet coefficients (Starck \& Pierre \cite{sta98},
Starck et al. \cite{mr1}). Each filtered image is then exposure
corrected and a mask map that includes bad pixels, CCD gaps and
non-exposed CCD regions (generally parts outside the field-of-view of
the telescope) are created. Wavelet-filtered, exposure-corrected
images for each instrument in a given energy band are added together
to form a compound (MOS1+MOS2+PN) single band image to be used in the
first stage of the detection procedure.

\subsection{Source detection}
Clusters of galaxies are extended sources in X-ray images. Their
detection and correct classification is not trivial because of various
peculiarities of the X-ray observations: Poisson noise regime, varying
PSF as a function of the off-axis angle and the energy, the vignetting
effect, and the geometry of the detector (CCD gaps). We use images in
the [0.5-2] keV energy band which is well suited for clusters and
groups (Scharf \cite{sch02}). The detection procedure is based on the
prescription of Valtchanov et al.  (\cite{val01}) and has three
stages: wavelet filtering (see the previous section), detection and
measurements. The wavelet filtered image is fed to {\tt SExtractor}
(Bertin \& Arnouts \cite{sex}) for detection and characterisation. The
classification to extended (clusters) and point-like sources (AGNs or
QSOs) is done by using three parameters: the half-light radius, the
FWHM from a Gaussian fit to the source and {\tt SExtractor} stellarity
index adapted for the X-ray observations. The classification task is
complicated by the fact that distant and faint clusters are not too
different from a PSF. Moreover, at greater off-axis angles the PSF
shape can be quite distorted, although the half-energy width does not
change significantly in the $[0.5-2]$ keV energy band. This shape
distortion can lead to a wrong classification and that is why we have
constrained the cluster detection up to off-axis distances not greater
than $12\arcmin$. This strategy gives very good results using
simulated XMM images and can be used for selecting cluster candidates
with an excellent success rate (Valtchanov et al.  \cite{val01}).

In total, from the first 3.5 deg$^2$ (30 XMM pointings) of XMM-LSS
observations, we find some 55 X-ray cluster candidates, which
corresponds to $\sim 15$ clusters per sq.deg., in good agreement with
the cosmological predictions (see e.g. Refregier et al. \cite{ref02}).

\section{Optical data}
\label{sec:opt}

\subsection{Optical imaging}

Deep images from the CFH12k camera on the Canada-France-Hawaii
Telescope (CFHT) in BVRI from the VIRMOS-VLT Deep Survey (VVDS: Le
F\`evre et al. \cite{virmos1}, McCracken et al. \cite{virmos2}) were
used for optical identifications.  The observations were processed by
the Terapix team\footnote{http://terapix.iap.fr} to produce
astrometrically and photometrically calibrated images and to create
object catalogs by means of {\tt SExtractor}. The definition and
magnitude limits of the VVDS may be found in the web pages of the
VIRMOS consortium\footnote{http://www.astrsp-mrs.fr/virmos/vvds.htm}.
The X-ray cluster candidates were assigned to {\em NEAR} ($z<0.5$) and
{\em MID} ($0.5<z<1$) samples by visual inspection of the optical
images and also using photometric redshift estimates. We assign all
the confident extended sources, that have no obvious optical
counterpart in the VVDS images, in the {\em DIST} ($z>1$) class for
subsequent NIR follow-up. The visual scan of the data was also
indispensable for removal of spurious extended sources introduced by
instrumental effects. In addition to the CFH12k data, we obtained
observations at the CTIO 4m telescope in R and z' which allowed us to
search for a red sequence of galaxies at a given X-ray position
(Andreon et al. \cite{and04}).

\subsection{Spectroscopic target selection}

Our driving objective was to confirm the clusters and to measure their
redshifts. We chose the ESO-VLT/FORS2 instrument in MXU multi-object
spectroscopy mode because of the liberty to place large number of
slits with different sizes and orientations and so to optimise the
target selection, especially in regions of high galaxy density.  Using
{\tt fims} (the FORS Instrumental Mask Simulator tool), we placed
about 30 slits on average in each mask, possibly placing inclined and
longer ones. With the remaining available slits we randomly sampled
the X-ray point-like population presented in the same area or just
filled the mask with objects from the field, when there were no colour
selected or X-ray objects.

We have developed a visualisation procedure to facilitate the
selection of spectroscopic targets, optimising the chances to pick up
cluster member galaxies. It uses all the available information in the
field, combining optical (one band or pseudo-colour VRI image) and
X-ray images, colour-magnitude and colour-colour diagrams, and
photometric redshift peaks with the corresponding probabilities.  In
this ``multi-parametric space'', first we have taken the objects above
the photometric limit, imposed by exposure time and observational
strategy constraints. Secondly, we have looked for objects with a
plausible colour (V$-$I or R$-$I) for ellipticals at the estimated
photometric redshift and also used the pseudo-colour images for all
candidates for which we have V, R and I observations.  The utility of
colour information is justified as, out to redshift of unity, the
major population of cluster cores is made of ellipticals with an old
stellar population (e.g. Dressler et al. \cite{dre97}, Postman et al.
\cite{pos98}). Moreover, cluster ellipticals at high redshift show-up
as an overdensity of red objects in VRI composite images, proving
useful in the target selection process. Examples of RGB images for
some of the candidate clusters can be found in the web pages of the
XMM-LSS
consortium\footnote{http://vela.astro.ulg.ac.be/themes/spatial/xmm/LSS}.

\subsection{Spectroscopic observations}

The ESO-VLT/FORS2 spectroscopic observations in MXU mode were
performed on 9-12 October 2002.  For the 3 allocated nights in
``visitor'' mode we selected some 12 candidates from the {\em NEAR}
and {\em MID} samples. {\em NEAR} clusters were included in order to
provide targets in case of a poor weather and will be presented
elsewhere (Willis et al., in preparation). The observing log and object
designations for the five {\em MID} clusters pertaining to this paper are
shown in Table~\ref{tab:obs}.

We have used the holographic grism 600RI+19 with the GG435 blocking
filter that gives a good response from about 6000 to 8500 \AA\ and a
dispersion of 1.64 \AA/pixel with the standard instrument resolution.
The spectral resolution for slits of $1.4\arcsec$ width as measured
from the arc or sky lines was $\approx 6$ \AA\ FWHM. The high
resolution of the grism ($\lambda/\Delta\lambda=1000$) makes
subsequent sky subtraction easier.

\begin{table}
  \caption[ ]{ESO-VLT observing log. The exposure time is the
number of masks times the time, and each mask is split into two
exposures.} 
  \begin{tabular}{llll}
    \hline
    \multicolumn{1}{c}{ID$^a$} &
    \multicolumn{1}{c}{Cluster} &
    \multicolumn{1}{c}{Seeing}   &
    \multicolumn{1}{c}{Exposure} \\
    & & \multicolumn{1}{c}{arcsec} &
    \multicolumn{1}{c}{min.}\\
    \hline\hline
    \multicolumn{4}{c}{09-Oct-2002} \\
     001 & XLSS J022457.1$-$034853 & 1.0-1.2 & 2x60 \\
     005 & XLSS J022709.7$-$041805 & 0.7-1.1 & 2x90 \\
    \multicolumn{4}{c}{10-Oct-2002} \\
     004 & XLSS J022530.2$-$050713 & 0.8-2.7 & 90+60 \\
    \multicolumn{4}{c}{10-Oct-2002} \\
    002 & XLSS J022532.5$-$035510 &  0.7-1.2 & 2x60 \\
    003 & XLSS J022738.2$-$031757 &  0.7-1.1 & 2x60 \\
    \hline
  \end{tabular}

\medskip
  $^a$ - In the following, all clusters are referred to via the
  reference XLSSC plus the identification number, e.g. XLSSC 001, etc.
  \label{tab:obs}
\end{table}

For each mask we split the total observing time into two exposures in
order to remove cosmic ray contaminations. We have used the {\tt IRAF}
task {\tt imcombine} with the {\tt crreject} algorithm -- appropriate
for rejecting cosmic ray events even with two images. The spectral
reduction was performed using the standard tools from {\tt IRAF}: zero
level exposure, flat-fielding, cosmic ray removal, aperture selection
and sky subtraction using the {\tt apextract} package. For wavelength
calibration, we have used He-Ne-Ar lamp exposures observed at the end
of the corresponding night. The wavelength calibration was performed
with a Chebyshev polynomial of 3rd degree and the residuals were kept
inside $\pm1.0$ \AA\ with an rms scatter $\sim 0.3-0.5$ \AA. A
standard star was observed in the beginning and at the end of each
night with the same instrumental configuration in order to remove the
instrumental response and therefore to transform the spectra to
relative flux units.

\section{Results}

\label{sec:results}

\subsection{Redshifts}

The sky subtracted, wavelength and flux calibrated spectra were used
to derive the redshifts. First we obtained a redshift estimate based
on the CaII H+K doublet, when present, and then we used the
cross-correlation technique as implemented in the {\tt RVSAO} package
(Kurtz \& Mink \cite{rvsao}) using an elliptical galaxy template
spectrum (Kinney et al. \cite{kin}).  To improve the cross-correlation
signal, we have masked the wavelength ranges of the strongest sky
lines, where residuals from the sky subtraction could occur, and also
the region of a strong atmospheric absorption at 7750-7800 \AA.

\begin{figure*}

  \centerline{
    \includegraphics[width=8cm]{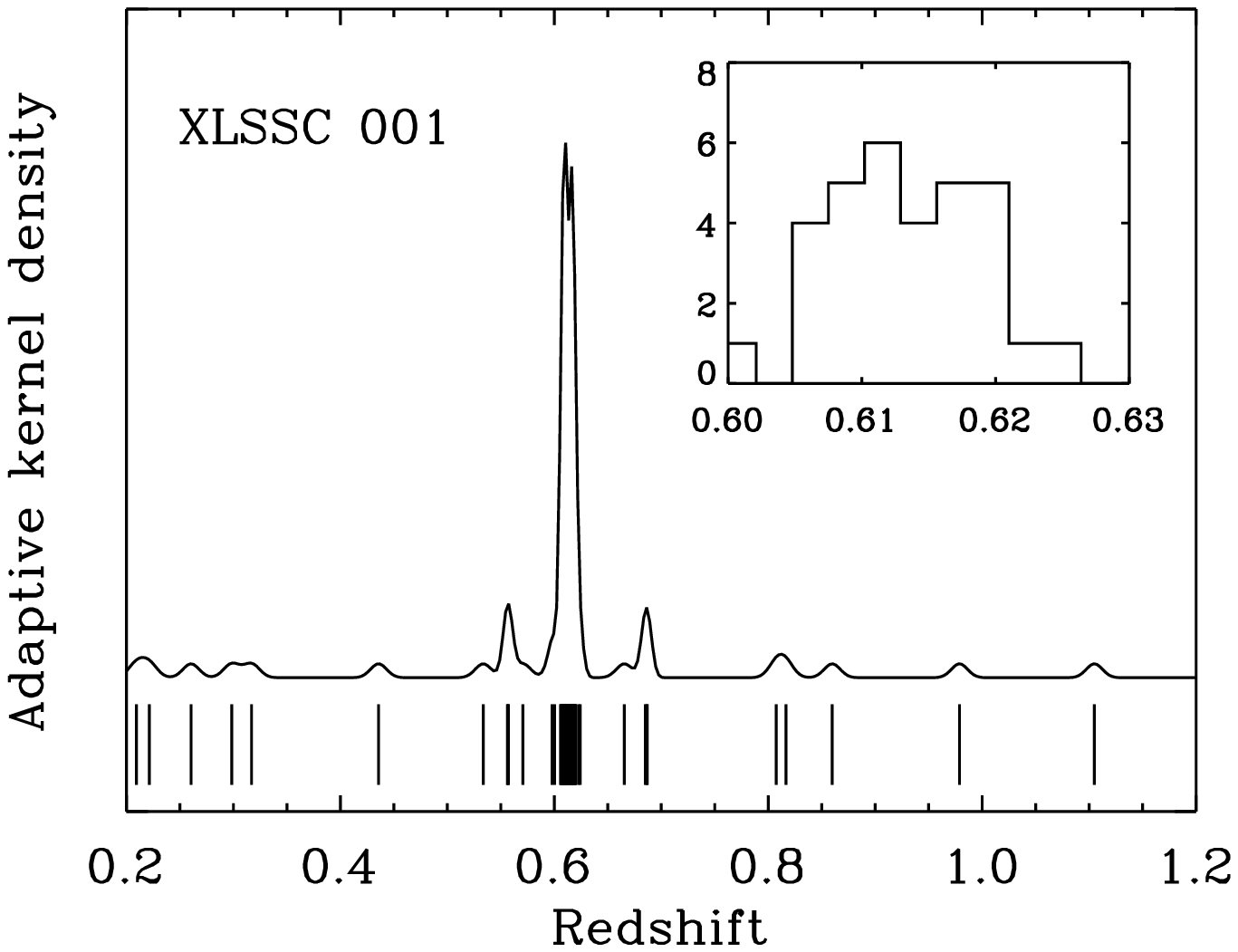} \hfill
    \includegraphics[width=8cm]{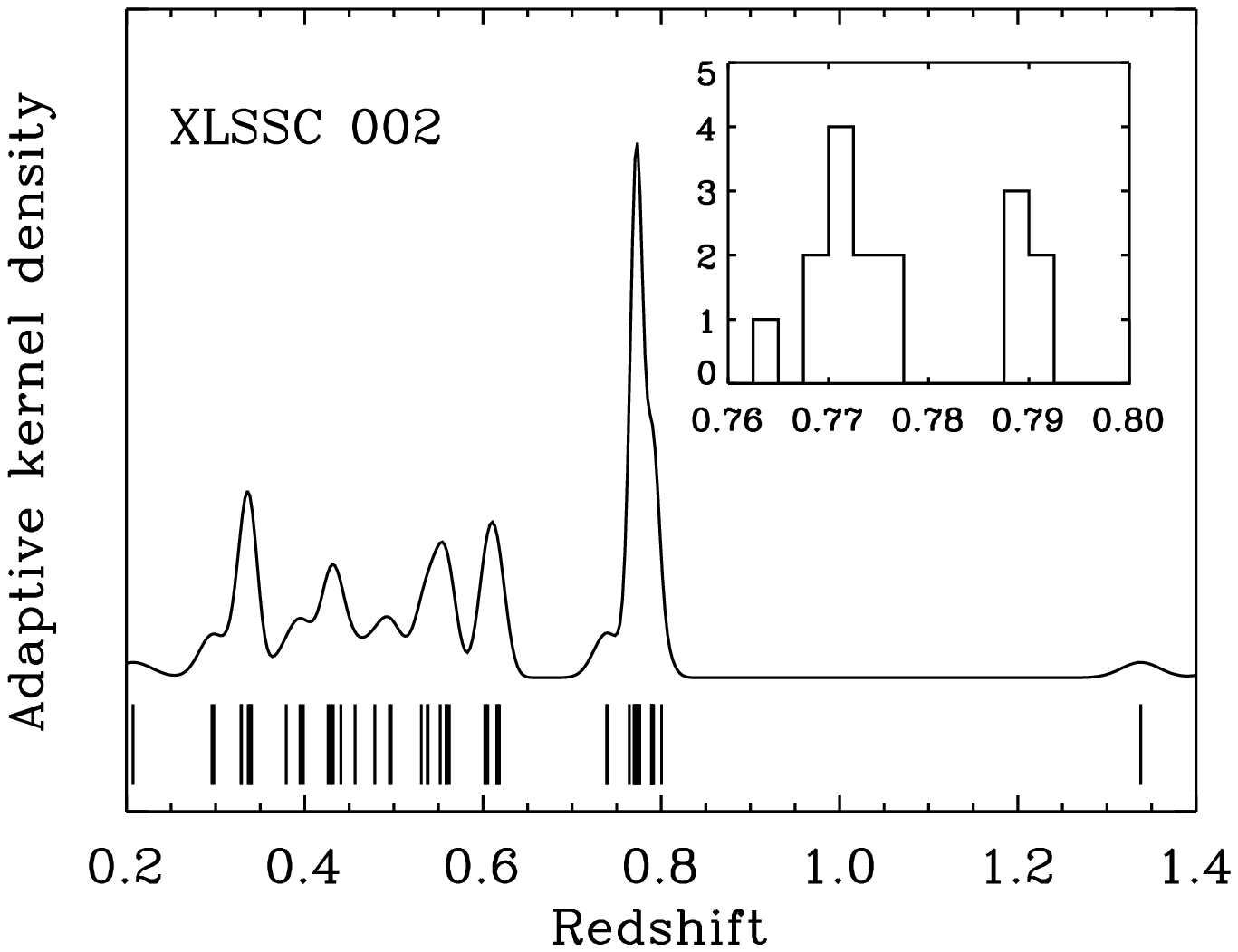}
  }

  \centerline{
    \includegraphics[width=8cm]{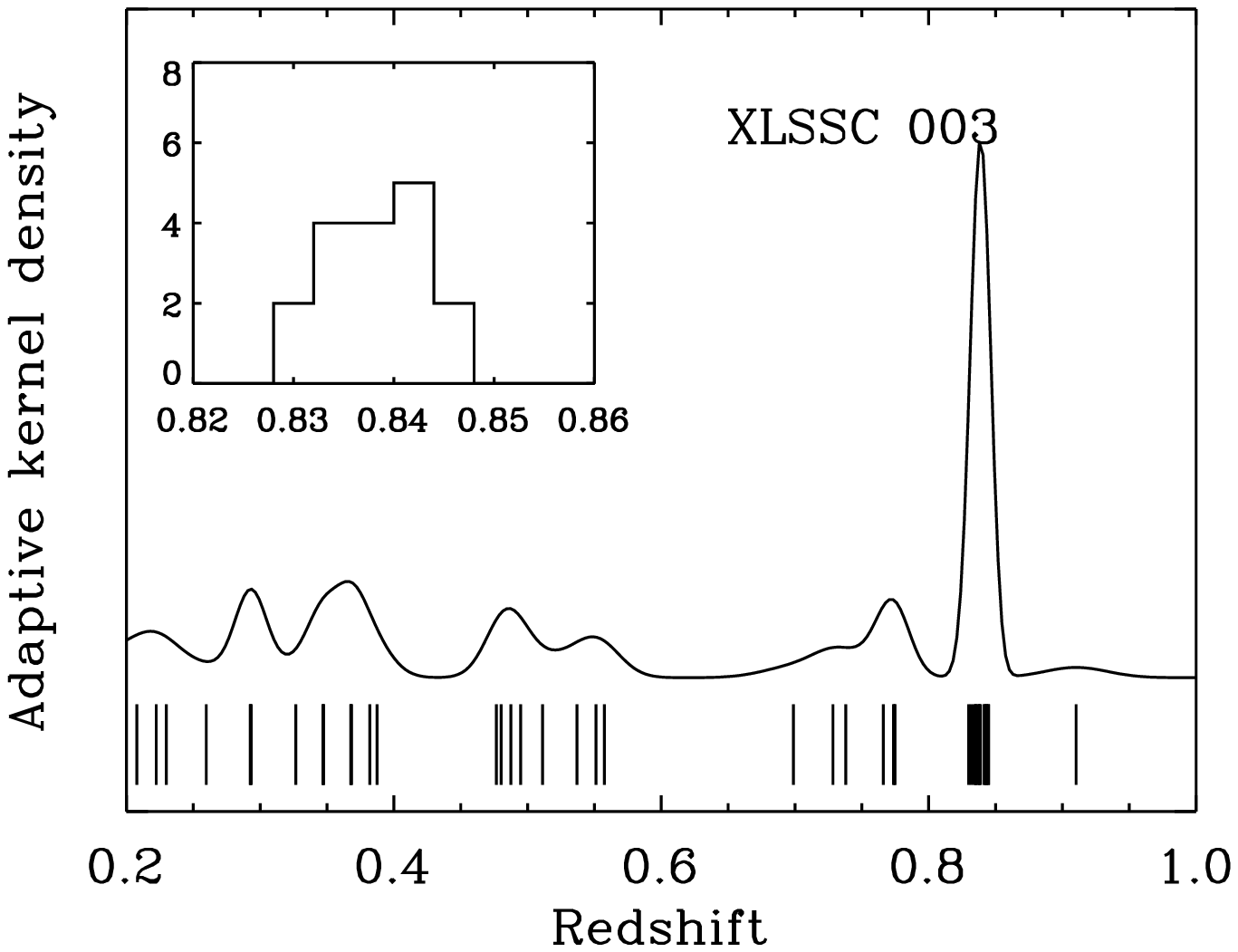} \hfill
    \includegraphics[width=8cm]{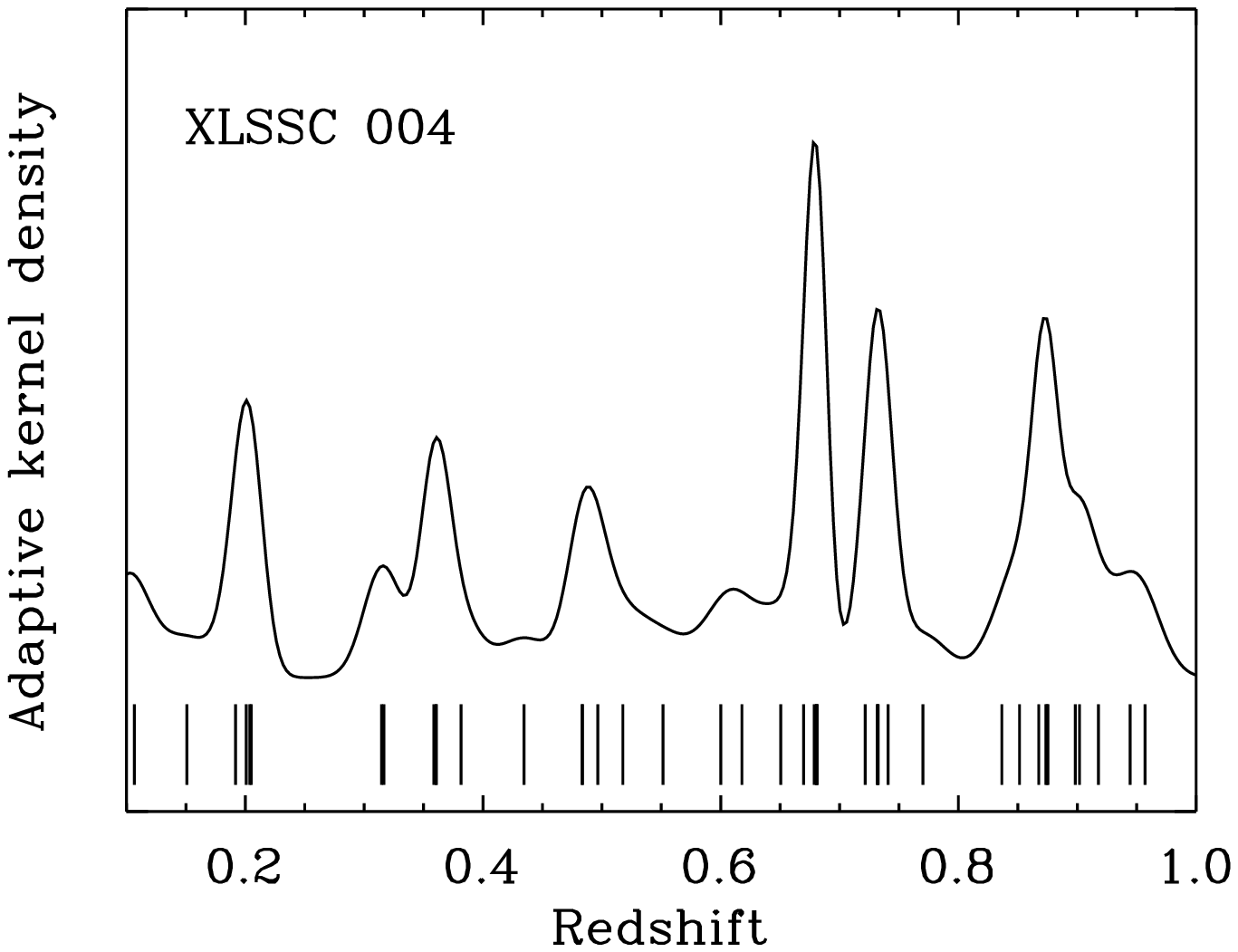}
  }

\centerline{ \includegraphics[width=8cm]{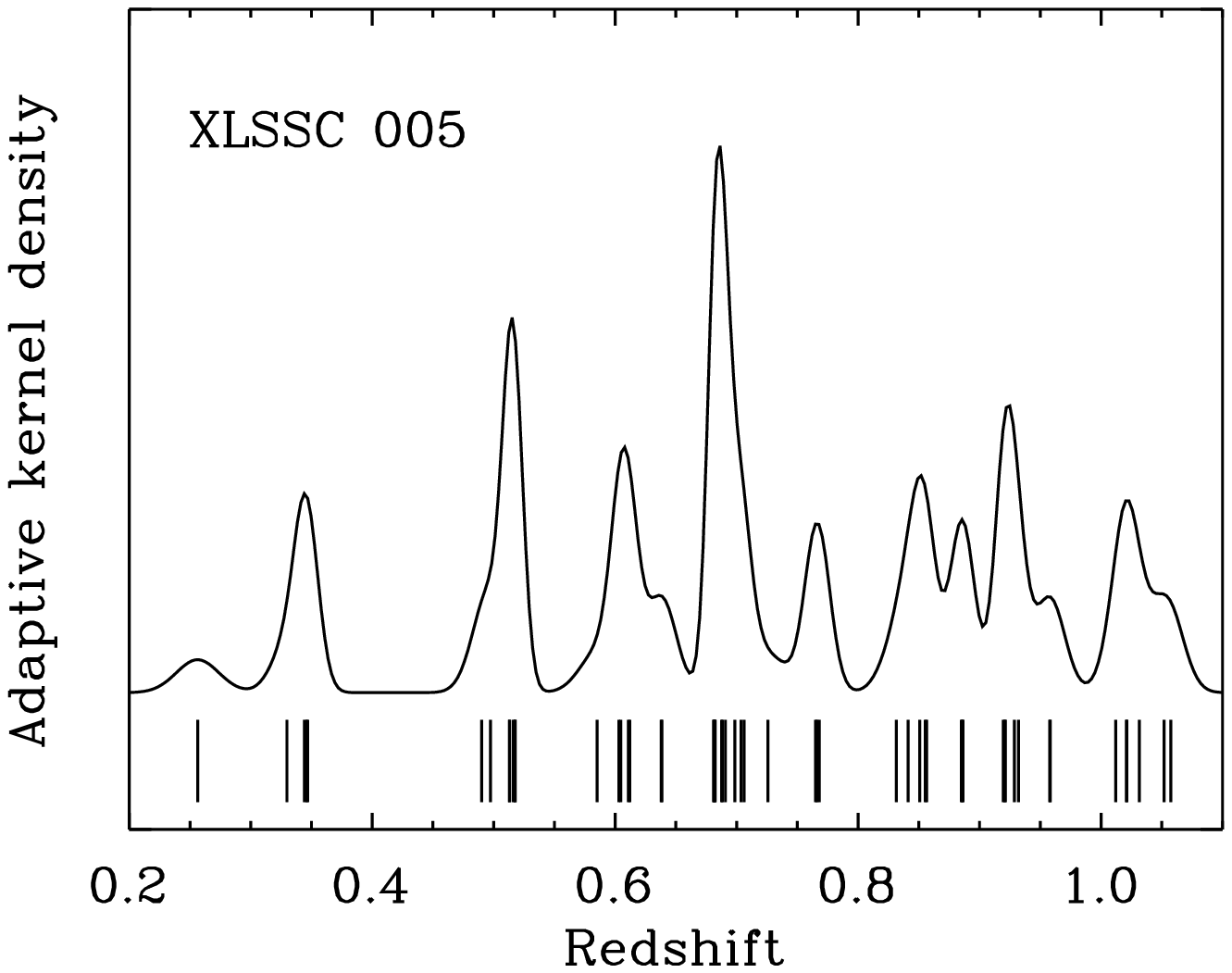}}

  \caption[]{Redshift distributions of the spectroscopically observed
  objects for all five clusters. The adaptive kernel density
  estimation (Silverman \cite{sil86}) is shown as a continuous line
  over the redshift measurements. In cases where an overdensity in
  redshift space corresponds to a spatial correlation we show a
  histogram of the zoomed redshift region: for \clsA\ from 0.6 to
  0.63, bin size 0.0027; \clsB: $0.76<z<0.80$, bin size 0.0025; \clsC:
  $0.82<z<0.86$, bin size 0.004.}  \label{fig:redshifts}

\end{figure*}

The redshift distributions in each field are shown in
Fig.~\ref{fig:redshifts}. To illustrate the redshift space
overdensities we have applied an adaptive kernel smoothing over the
redshifts (e.g. Silverman \cite{sil86}, Pisani \cite{pis93}). There is
both a clear redshift grouping and a spatial grouping associated with
the X-ray emission in \clsA, \clsB\ and \clsC. For these three cases
only we show zoomed in an inset a redshift histogram of the
overdensity with a fixed bin size. The derived mean redshift and the
velocity dispersion, by means of bi-weighted estimators of location
and scale (Beers et al. \cite{rostat}), are shown in
Table~\ref{tab:spec}; the quoted errors are $1\sigma$ bias-accelerated
bootstrap errors (Efron \& Tibshirani \cite{efr86}).

For \clsD\ and \clsE, the peaks in the redshift distribution over the
whole FORS field do not correspond to any significant spatial
clustering, and no peak can be unambiguously associated with the
extended X-ray source. Nevertheless, we show in Table~\ref{tab:spec}
the most plausible redshift based on measurements of galaxies
spatially coincident with the cluster X-ray emission (see
Figs.~\ref{fig:clsd:image} and \ref{fig:clse:image}).

\begin{table}

  \caption{Spectroscopic redshift measurements. N$_{gal}$ shows the
  total number of galaxies used to derive the cluster redshift. The
  number in parenthesis shows the number of galaxies with [OII]
  3727\AA\ emission line. See the text for details on \clsD\ and
  \clsE\ redshifts.}

  \label{tab:spec}
  \begin{tabular}{lllr}
    \hline
    \multicolumn{1}{c}{Name} &
    \multicolumn{1}{c}{Redshift} &
    \multicolumn{1}{c}{$\sigma_v$} &
    \multicolumn{1}{c}{N$_{gal}$} \\
    & & \multicolumn{1}{c}{km/s}\\
    \hline\\
    \clsA\ & $0.6128^{+0.0006}_{-0.0006}$ & $867^{+80}_{-60}$ & 29(9) \\[12pt]
    \clsB$^{*}$ & $0.7722^{+0.0001}_{-0.0003}$ & $524^{+267}_{-116}$ &
    11(5) \\[12pt]
    \clsC\ & $0.8387^{+0.0007}_{-0.0007}$ & $780^{+137}_{-75}$ & 17(5) \\[12pt]
    \clsD\ & $\sim 0.87$ & --- & 4(1) \\
    \clsE & $\sim 1.0$ & --- &  7(5)\\
    \hline
  \end{tabular}

  \medskip

  (*) -- taking only the objects having redshifts in [$0.76,0.78$].

\end{table}

\subsection{X-ray analysis}

Deriving detailed cluster physical characteristics from the X-ray
observations is not the main objective of XMM-LSS. Indeed, with
exposure times of the order of 10 ks only a small fraction of clusters
are expected to have enough photon statistics to allow reliable
measurements of the mean temperature, or to distinguish possible AGN
contamination.  Nevertheless, we have derived the temperature and
luminosity for the clusters pertaining to this paper and, in some
cases, these parameters were well constrained. These results however
must be taken with caution as in some cases possible AGN contamination
cannot be ruled out.

For each cluster an X-ray spectrum was extracted from a region large
enough to include the cluster emission. A background spectrum was
taken from an adjacent annulus. We removed in advance the contribution
of all other sources within the cluster and background regions. A
photon redistribution matrix (RMF) and ancillary region file (ARF)
were created using {\tt XMMSAS:rmfgen/arfgen}, including corrections
for bad pixels and detector geometry. Finally the spectra from the
three instruments MOS1, MOS2 and PN were regrouped to have at least 25
counts per bin. The extracted spectra were used to derive the observed
characteristics shown in Table~\ref{tab:x1} without any recourse to a
reference model.

\begin{table*}

  \caption{X-ray parameters. ``Region'' is the extraction radius in
  arcmin and in comoving Mpc in brackets. ``$R_{50}$'' is the object's
  half-light radius in arcsec in the [0.5-2] keV band, note that the
  PSF half-light radius at $10\arcmin$ off-axis and at 1 keV is
  $9\arcsec$. The ``Exposure'' is the average of the weighted live 
  time events in the extraction region.}  

  \label{tab:x1}
  \begin{tabular}{lllllll}
    \hline
    \multicolumn{1}{c}{Name} &
    \multicolumn{1}{c}{Redshift} &
    \multicolumn{1}{c}{Region} &
    \multicolumn{1}{c}{$R_{50}$} &
    \multicolumn{1}{c}{off-axis} &
    \multicolumn{1}{c}{Counts} &
    \multicolumn{1}{c}{Exposure}\\
    & & & & &
    \multicolumn{1}{c}{[0.2-10] keV} &
    \multicolumn{1}{c}{{\tiny MOS1+MOS2+PN}}\\
    & &
    \multicolumn{1}{c}{{\tiny arcmin(Mpc)}} &
    \multicolumn{1}{c}{arcsec} &
    \multicolumn{1}{c}{arcmin} &
    \multicolumn{1}{c}{{\tiny ALL(MOS1+MOS2+PN)}} &
    \multicolumn{1}{c}{ks} \\
    \hline\\
    \clsA\ & 0.6128 & 1.7(0.78)  & 32.2 & 5.8 & 1812(472+467+873) & 2x13.3+7.9 \\
    \clsB\ & 0.7722 & 1.1(0.60)  & 16.6 & 9.9 & 633(177+152+304)  & 2x13.3+7.9 \\
    \clsC\ & 0.8378 & 1.1(0.65)  & 23.6 & 6.0 & 575(136+128+311)  & 2x11.5+8.3  \\
    \clsD\ & 0.87   & 0.5(0.27)  & ---  & 3.8 & 233(40+83+110)    & 2x20.8+16.6  \\
    \clsE\ & 1.0    & 0.83(0.56) & 21.1 & 8.5 & 353(50+76+227)    & 24.5+24.9+21.5 \\
    \hline
  \end{tabular}
\end{table*}

To get the global cluster X-ray characteristics, the binned spectra in
each instrument were fitted to a {\tt mekal} model of thermal plasma
emission with photo-electric absorption using {\tt XSPEC} (Arnaud
\cite{xspec}, also see the {\tt XSPEC}
manual\footnote{http://heasarc.gsfc.nasa.gov/docs/xanadu/xspec/manual/}
for the {\tt mekal} model and corresponding references). The energy
range used in the fit is $[0.3-10]$ keV for MOS instruments, but for
PN we used $[0.3-7.5]$ keV in order to avoid an instrumental emission
feature at $8-9$ keV. We have kept only two free parameters in the
fitting: the temperature and the normalisation.  The mean Galactic
absorption of $N_H=2.5\times 10^{20}$ cm$^{-2}$ (Dickey \& Lockman
\cite{nh}), the metal abundance of $Z=0.3Z_{\sun}$ and the redshift
were fixed. For parameter estimation, we used the Cash statistic,
modified to allow background subtraction (see the {\tt XSPEC} manual).
The results are shown in Table~\ref{tab:x2}.  The errors on the
temperature are $1\sigma$, the errors for the flux $F_X$ and the
luminosity $L_X$ were calculated varying the normalisation parameter
in its $1\sigma$ confidence interval, while the temperature was kept
fixed at its best fit value. The bolometric X-ray luminosity $L_X$ was
calculated using the energy range $[0.01-30]$ keV, constrained only by
the instrument response matrices (generally from $\sim 0.05$ to $\sim
20$ keV rest frame).

For \clsD\ and \clsE\ photon statistics are insufficient to allow any
plausible parameter estimation so we have estimated the flux and the
luminosity by fixing the temperature at 4 keV. The uncertainties in
this case came from varying only the normalisation and thus can be
quite underestimated.

The cluster masses were estimated from the local $L_X-M$ relation
(Reiprich \& B\"ohringer
\cite{rei02}) assuming no evolution. 
The errors on the mass, reported in Table~\ref{tab:x2}, only
reflect the uncertainty on the luminosities and so are significantly
underestimated, given the large (up to an order of magnitude!) scatter
in the local $L_X-M$ relation.

\begin{figure}

  \centerline{
    \vbox{
    \includegraphics[height=8cm,angle=270]{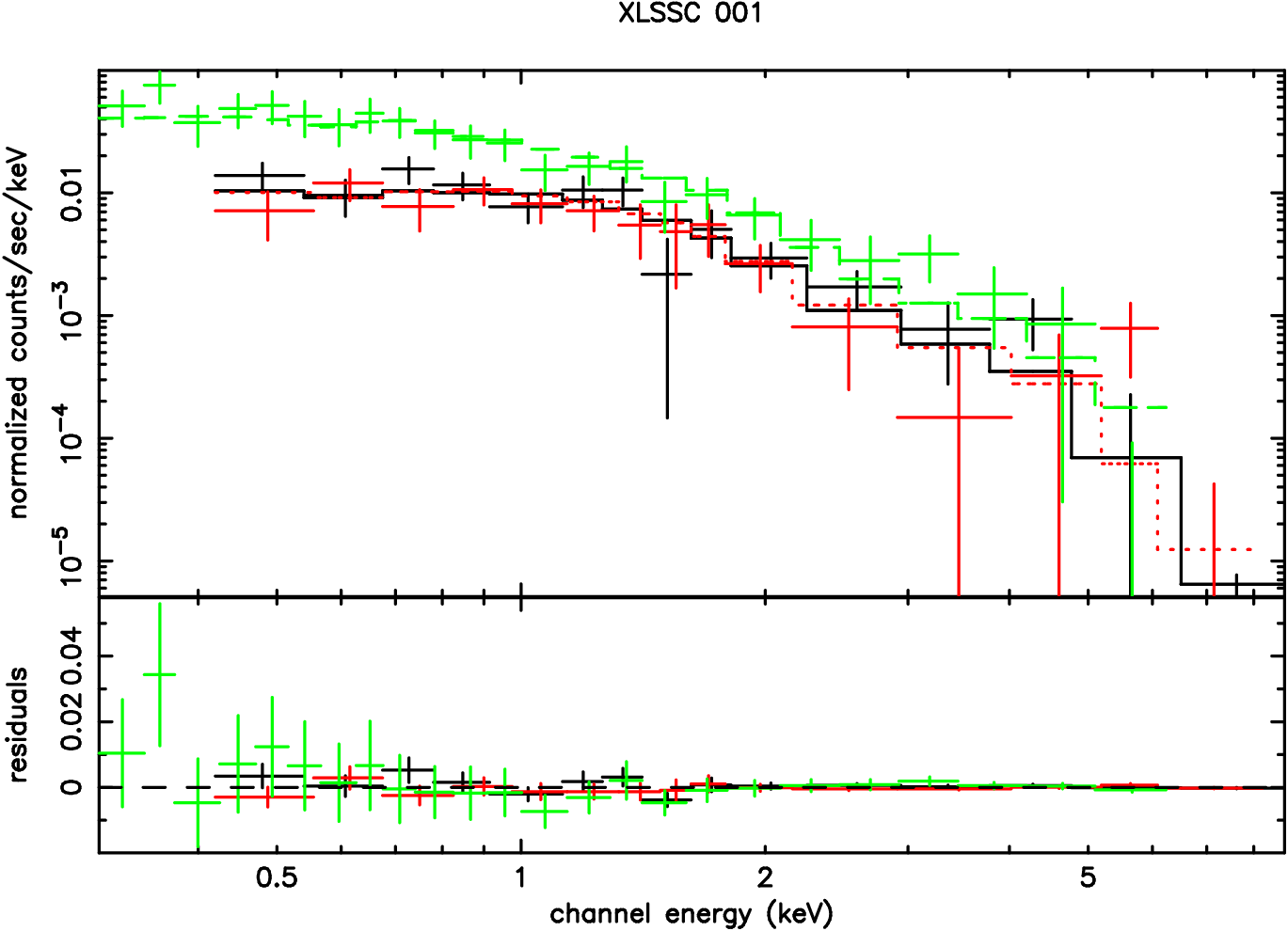} \\
    \includegraphics[height=8cm,angle=270]{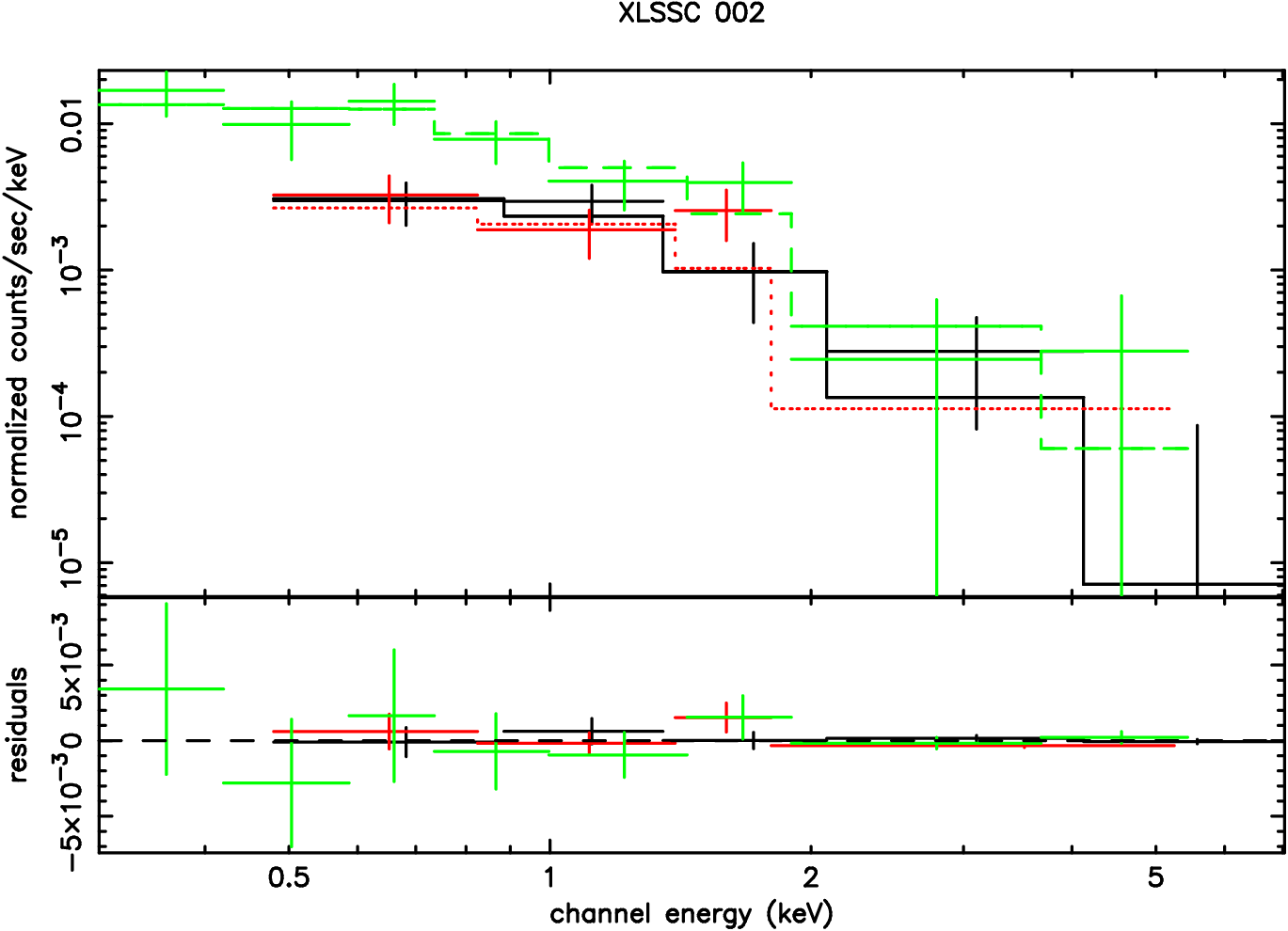}\\
    \includegraphics[height=8cm,angle=270]{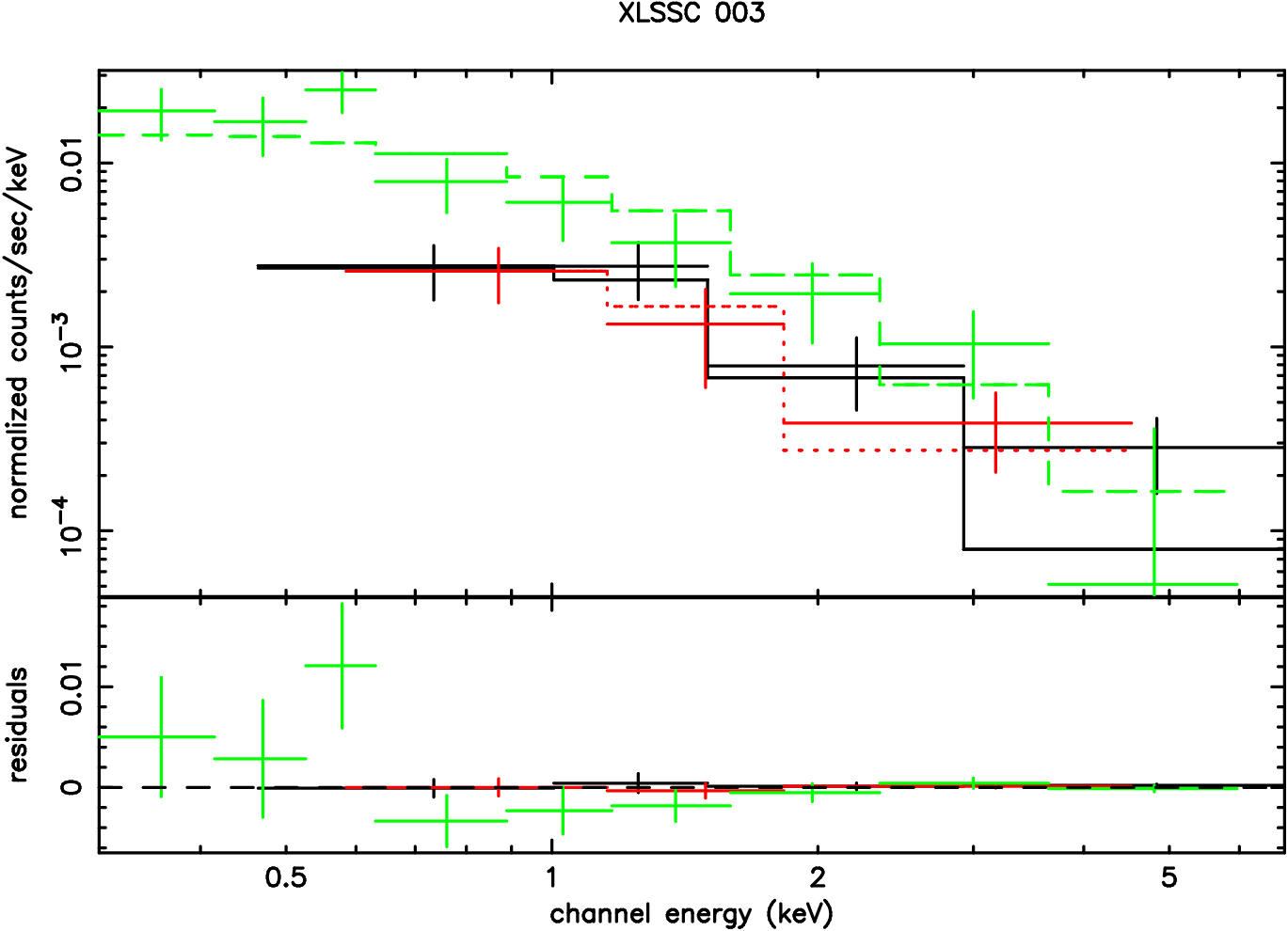}}}

  \caption[]{X-ray spectra, model fit and the corresponding residuals
  for MOS1 (in black, the continuous histogram shows the model fit),
  MOS2 (in red, the dotted histogram is the model fit) and PN (in
  green, the model fit is the dashed histogram). The spectra were
  initially regrouped to have at least 25 photons per energy bin.}

\label{fig:xspec}
\end{figure}

The X-ray spectra, {\tt mekal} fit and the corresponding residuals for
\clsA, \clsB\ and \clsC\ are shown in Fig.~\ref{fig:xspec} and the
derived model characteristics are given in Table~\ref{tab:x2}.

\begin{table}
  \caption{Derived, model dependent, cluster X-ray
  characteristics. The data presented for \clsD\ and \clsE\ is for
  fixed temperature of 4 keV and the uncertainties are probably
  underestimated. The reported masses should be treated with
  extreme caution as the scatter in the local $L_X-M$ relation is
  large. See text for details.}

  \label{tab:x2}
  \begin{tabular}{lllllll}
    \hline
    \multicolumn{1}{c}{Name} &
    \multicolumn{1}{c}{Redshift} &
    \multicolumn{1}{c}{$kT$} &
    \multicolumn{1}{c}{$F_X^{a}$} &
    \multicolumn{1}{c}{$L_X^{b}$ } &
    \multicolumn{1}{c}{$M^{c}$} \\
    &  &  \multicolumn{1}{c}{keV} \\
    \hline\\
    \clsA\ & 0.6128 & $3.4^{+1.1}_{-0.8}$ & $8.9^{+1.0}_{-0.7}$ &
    $3.5^{+0.3}_{-0.3}$ & $4.4^{+0.2}_{-0.2}$ \\[12pt]
    \clsB\ & 0.7722 & $2.2^{+1.3}_{-0.6}$ & $2.6^{+0.7}_{-0.4}$ &
    $1.7^{+0.4}_{-0.3}$ & $2.9^{+0.5}_{-0.5}$ \\[12pt]
    \clsC\ & 0.8378 & $4.1^{+4.7}_{-1.8}$ & $4.4^{+1.0}_{-0.8}$ &
    $3.7^{+0.9}_{-0.6}$ & $4.3^{+0.5}_{-0.4}$ \\[12pt]
    \clsD  & 0.87   & 4 & $0.23^{+0.2}_{-0.2}$ & $0.21^{+0.1}_{-0.1}$ &
        $0.8^{+0.9}_{-0.9}$\\[12pt]
    \clsE  & 1.0    & 4 & $1.03^{+0.3}_{-0.2}$ & $1.32^{+0.3}_{-0.3}$ &
        $2.0^{+0.3}_{-0.3}$\\
    \hline
  \end{tabular}
  \medskip

  $^{a}$ -- the unabsorbed flux $F_X$ in the [0.5-2] keV band and in
  units of $10^{-14}$ \flux.

  $^{b}$ -- the luminosity $L_X$ is pseudo-bolometric ([0.01-30] keV)
  and in units of $10^{44}$ erg s$^{-1}$.

  $^{c}$ -- the total mass $M$ in units of $10^{14}\,M_{\sun}$, based
  on the local $L_X-M$ relation.

\end{table}

\subsection{Scaling relations}

\begin{figure}
\vbox{

\includegraphics[width=8cm]{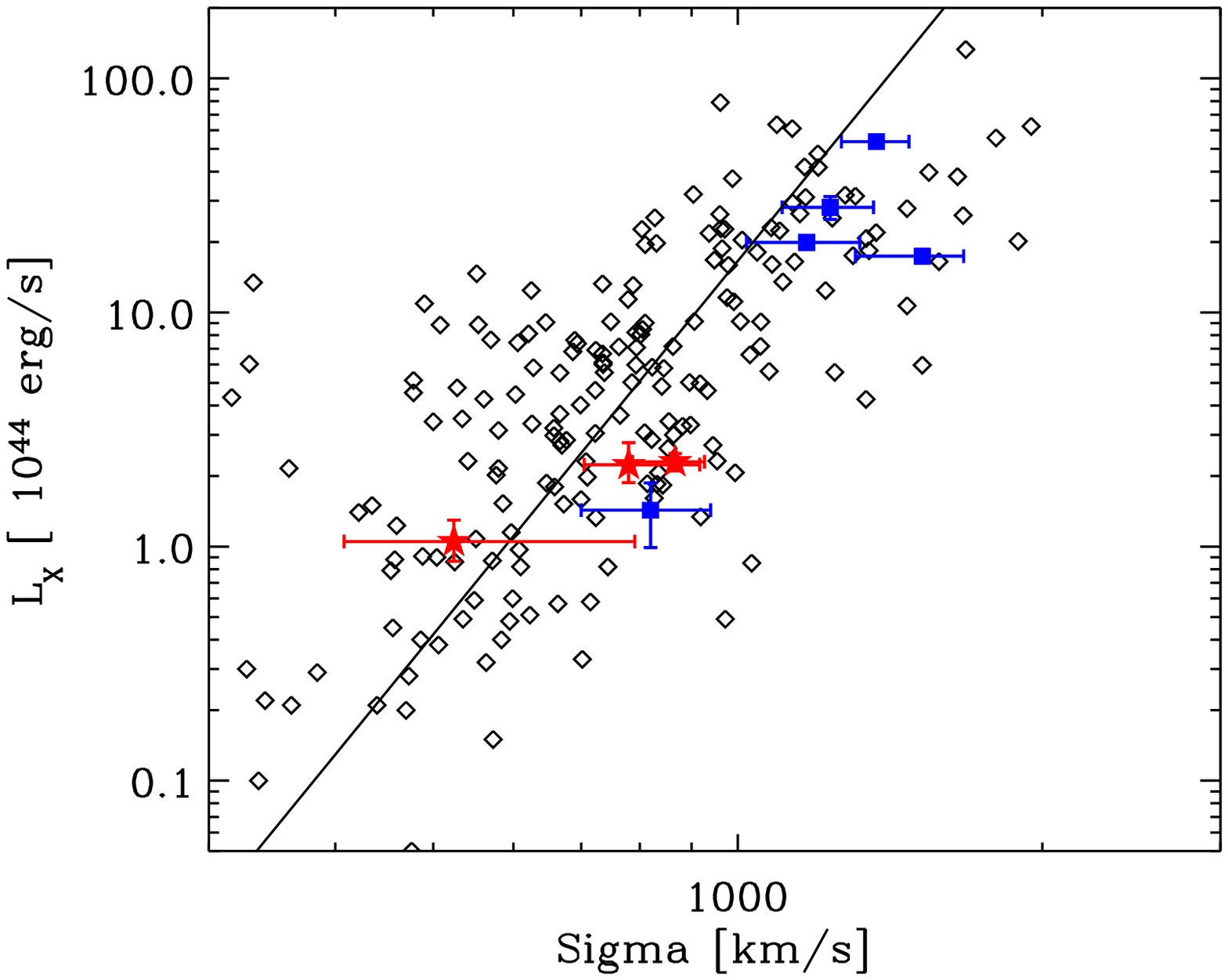}

\includegraphics[width=8cm]{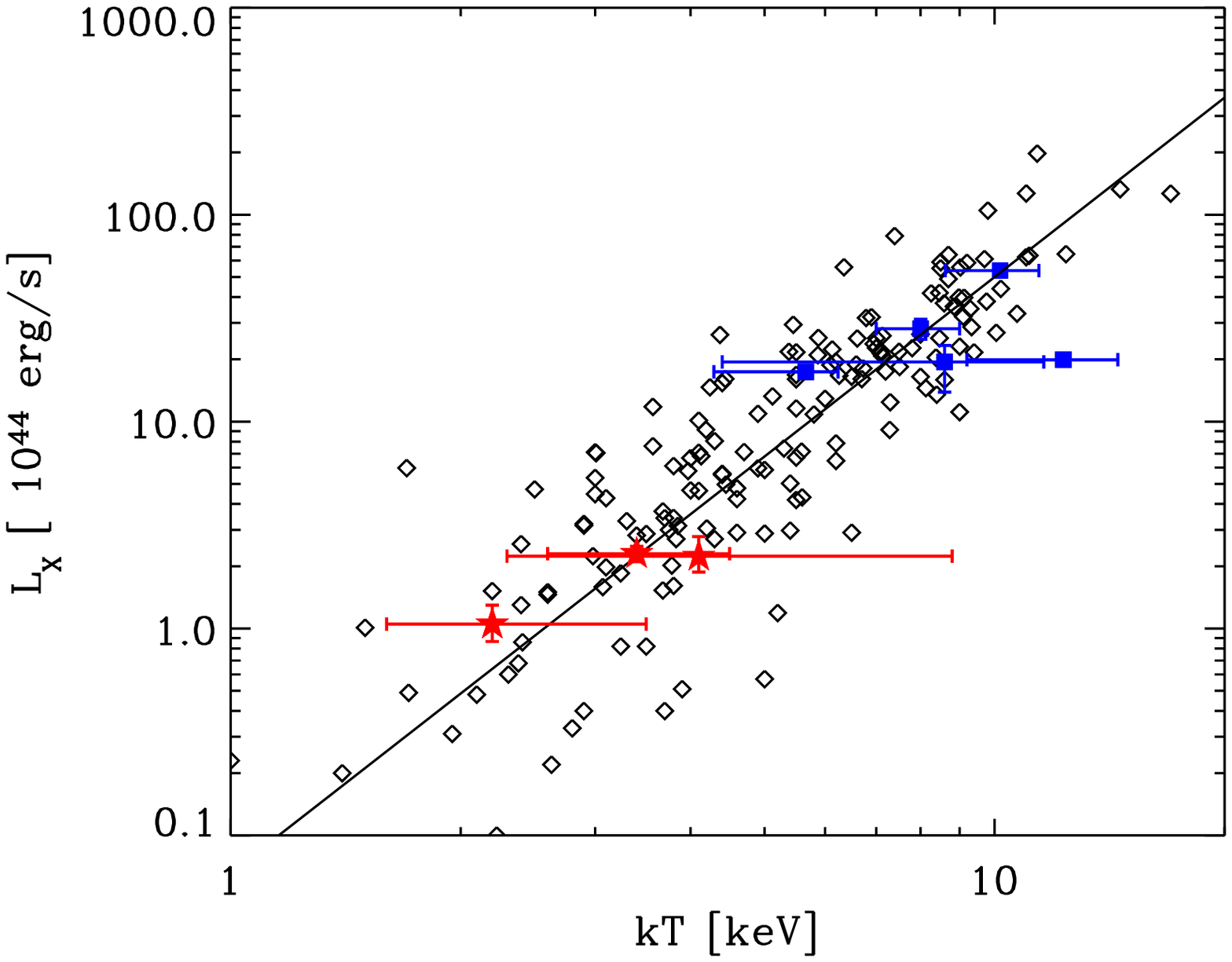}

\includegraphics[width=8cm]{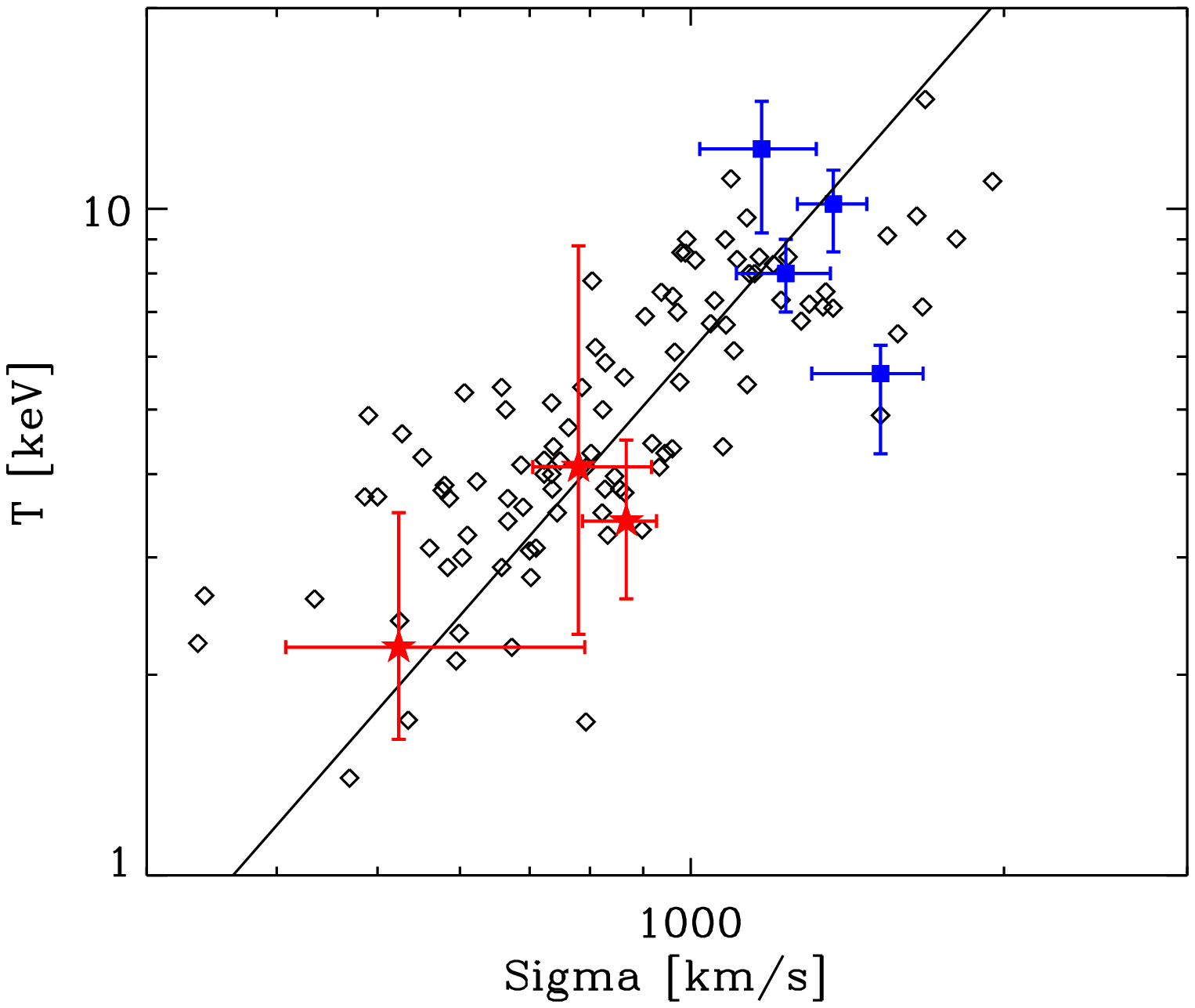}}

  \caption[]{The scaling relations: luminosity-velocity dispersion
  ($L_X-\sigma_v$, upper panel), luminosity-temperature ($L_X-T$,
  middle panel) and velocity dispersion-temperature (lower panel).
  The clusters from the compilation of Wu et al. (\cite{wu99}) are
  denoted as diamonds and those at $z\geq 0.5$ are shown as blue boxes
  with error bars. The best fit from Xue \& Wu (\cite{xue00}) for
  $L_X-\sigma_v$ and $\sigma_v-T$ and from Arnaud \& Evrard
  (\cite{arn99}) for $L_X-T$ are shown as lines. The objects
  pertaining to this paper are shown as filled red stars with the
  corresponding error bars. For this comparison only, their luminosity
  was transformed to Einstein-de Sitter cosmology with $H_0=50$ km
  s$^{-1}$ and $\Lambda=0$ ($q_0=0.5$). }

  \label{fig:scaling}
\end{figure}

The scalings between the bolometric luminosity, temperature and
velocity dispersion from the clusters in the compilation of Wu et al.
(\cite{wu99}) are shown in Fig.~\ref{fig:scaling}, together with the
objects pertaining to this paper. \clsA, \clsB\ and \clsC\ are in good
agreement with the scaling relations $L_X-\sigma_v$, $L_X-T$ and
$\sigma_v-T$ (Xue \& Wu \cite{xue00}, Arnaud \& Evrard \cite{arn99}),
although caution must be taken because of possible biases on the
temperature introduced by the insufficient photon statistics (see e.g.
Fairley et al. \cite{fai00}). What is important to note is that the
three clusters are at lower temperatures and luminosities than the
other known clusters at high redshift. In that sense we are starting
to reach moderately massive systems at high redshift, whilst from
ROSAT or EINSTEIN surveys the distant clusters are rather more
luminous and consequently more massive systems.

\section{Individual objects}
\label{sec:obj}

The optical/X-ray overlays for all candidates are shown in
Figs.\ref{fig:clsa:image}-\ref{fig:clse:image}. The optical images are
from CFHT/CFH12k camera in the I-band, the X-ray contours are from
wavelet filtered MOS1+MOS2+PN images (see Sec.~\ref{sec:x}). For
\clsA, \clsB\ and \clsC, the objects from the inset histograms of their
redshift distribution (Fig.~\ref{fig:redshifts}) are denoted as boxes
and triangles, the latter for the galaxies with [OII] 3727 \AA\
emission line. When there was no obvious redshift peak (\clsD\ and
\clsE) we indicate the redshift of each spectroscopically observed
galaxy.

Based on the morphology, redshift distribution and the visual
appearance, we define three broad cluster classes: relaxed, relaxing
and collapsing.

\subsection{Relaxed cluster: \clsC}
\label{sec:pre14}

\clsC\ (Fig.~\ref{fig:clsc:image}) has the appearance of a relaxed
cluster with velocity distribution close to a Gaussian. Normality
tests (Anderson-Darling and Shapiro-Wilk, see e.g. D'Agostino \&
Stephens \cite{dag86}) accept the null hypothesis of the distribution
being drawn from a normal distribution at 90\% confidence, but this
must be taken with caution as the number of objects is small. The
X-ray emission is slightly elongated and there is no evidence of
substructure.  The brightest cluster galaxy coincides with the X-ray
emission peak and is at rest with respect to the cluster mean
($z_{BCG}=0.8382$, $\Delta v=65$ km s$^{-1}$).

As can be seen from Tabs.~\ref{tab:x1} and \ref{tab:x2}, the low
photon statistics do not allow the temperature to be constrained,
although if we assume an equipartition between the galaxy velocities
and the gas, i.e. $kT = \mu m_p \sigma^{2}_v$, we derive a temperature
of $kT=3.9\pm1.5$ keV -- in excellent agreement with the best fit
value.

\subsection{Relaxing cluster: \clsA}
\label{sec:pre01}

Morphologically \clsA\ (Fig.~\ref{fig:clsa:image}) looks like present
day still relaxing clusters: there is an obvious centre formed by the
brightest cluster galaxies displaced at $\sim 10\arcsec$ from the
X-ray emission peak. The brightest cluster galaxy (BCG) has a typical
giant elliptical galaxy absorption spectrum and is at 750 km s$^{-1}$
with respect to the mean cluster redshift. The velocity
  distribution is quite broad, but the null hypothesis that the
  observed distribution is drawn from normal cannot be ruled out
  -- the Anderson-Darling statistics value
  is 0.527 which is less than the critical value 0.616 needed to reject the normality
  assumption at 90\% confidence level.

There are two other X-ray sources projected over the cluster
emission. One is the subclump to the north-east, which is most likely
associated with the cluster as there are three cluster members within
the X-ray contours (two of which are emission line galaxies) and no
obvious optical counterpart at the X-ray peak. Unfortunately, due to
instrumental problems, we could not get the optical spectrum of the
south-east point-like source.

The X-ray characteristics of the cluster are relatively well
constrained as we have about 1600 photons in the spectrum and the
cluster is of low temperature. The low luminosity of the cluster is in
agreement with its temperature and the velocity dispersion, as
compared to the scaling relations of local clusters
(Fig.~\ref{fig:scaling}).

\subsection{Collapsing cluster: \clsB}

\clsB\ (Fig.~\ref{fig:clsb:image}) resembles a point-like source in
X-rays, although its half-light radius is $R_{50} \approx17\arcsec$
(see Table \ref{tab:x1}, to be compared to the PSF half-energy radius
of $9\arcsec$) and the extension log-likelihood (as reported by the
XMM-SAS task {\tt emldetect}) is 7.4, corresponding to probability of
$6\times10^{-4}$ of being a point-like source. The 8 centrally located
galaxies form a crown-like figure, and five are in a very narrow
redshift range $0.771-0.775$ most likely forming a compact core.  The
brightest cluster galaxy is at redshift coinciding with the mean of
the centrally located galaxies ($z_{BCG}=0.7714$) and has a typical
giant elliptical galaxy absorption spectrum.

We have to note that there are galaxies at similar redshifts ($z
\approx 0.77$) all over the field as can be seen in
Fig.~\ref{fig:clsb:image}. None of the other X-ray sources present in
the field are cluster members.

The temperature is quite tightly constrained by the X-ray spectral fit
even though the photon statistics seems rather poor -- 480 photons in
the [0.2-10] keV band in total.  This is possible as the cluster is at
low temperature. The relations between cluster global characteristics,
$L_X,\ T$ and $\sigma_V$ (Fig.~\ref{fig:scaling}) are consistent with
those observed from local clusters.

There is a large fraction of emission-line galaxies in the centre (5
from 8) as identified by the presence of the [OII] 3727 \AA\ line.
This line is not broad in Type 1 AGN, and thus it is difficult to
assess the presence of an AGN with only this one emission feature --
most of the diagnostic line ratios that can be used involve lines
($H_{\beta}$, [OIII] 4959+5007 \AA, $H_{\alpha}$) outside the range of
the grism RI600 used in this spectroscopic run. Optical-to-X-ray flux
ratios from deep surveys suggest that AGN with magnitudes typical of
these galaxies have X-ray fluxes covering a wide range, between
$\approx$8x10$^{-16}$ and $\approx$8x10$^{-15}$ erg cm$^{-2}$ s$^{-1}$
(0.5-2 keV), with the mean being $\approx$2x$10^{-15}$ erg cm$^{-2}$
s$^{-1}$ (eg. Mainieri et al. \cite{mai02}). Even in the unlikely
event that all 5 emission line galaxies contain X-ray emitting AGN at
this mean flux, the total would only be $\approx$40\% of the measured
cluster flux. It is perhaps more likely that most are Butcher-Oemler
star-forming galaxies, implying an interestingly high star-forming
fraction (eg. Fairley et al. \cite{fai02} find blue fractions up to
40\% in low $L_X$ clusters at z$\sim$0.3).  Pending further studies of
the galaxy population, we conclude that some contamination by AGN may
be possible, but that it is unlikely that all the X-ray flux
originates in AGN.

Based on the morphology and the redshift distribution, the most
plausible interpretation is that the cluster is in its early stage of
formation: a compact core is already formed and the accretion of
matter from the nearby large-scale structure is underway -- many
galaxies in the field have redshifts around the redshift of the
central core.

\subsection{Other cases}

\underline{\em \clsD\/:} The cluster X-ray emission (see
Fig.~\ref{fig:clsd:image}) lies between two bright X-ray sources
identified as QSOs at $z = 1.12$ (the closest to the north-west, with
486 counts in the [0.5-5] keV band and V = 21.2, R = 20.9, I = 20.8)
and at $z = 1.19$ (to the south-east, with 378 counts and V = 18.5, R
= 18.5, I = 18.0). These two QSOs are located in diametrically opposed
directions at $0.9\arcmin$ and $1.2\arcmin$ from the centre of the
cluster X-ray emission, respectively. Such a configuration is very
unlikely to occur by chance.  Our estimated surface density is
approximately 30 X-ray sources per deg$^2$ over a field of 2.8 deg$^2$
in the XMM-LSS survey with detected counts greater than 300,
subsequently the probability of detecting two X-ray sources within
$2\arcmin$ from the cluster X-ray centre is about 1\%. Given that
these two quasars are located just behind a putative foreground
cluster at z = 0.9 (see below), gravitational lensing could possibly
affect the real positions and fluxes of these two background
objects. However, the hypothesis that these QSOs could be two images
of a single source lensed by a foreground cluster does not hold as
their optical spectra show a significant redshift difference $\Delta z
\simeq 0.07$. A more detailed discussion on this unusual association
will be presented elsewhere (Jean et al., in preparation).

We cannot confirm the reality of the cluster from the spectroscopic
observations only -- the brightest galaxy inside the X-ray emission is
at $z=0.868$ and there is another one at $15\arcsec$ to the south-east
at a similar redshift, $z=0.874$. Two more galaxies at north-west are
at similar redshift $0.875$, although outside the detected X-ray
emission.

\clsD\ is, however, independently detected on a scale of 1 arcmin diameter
(marked as a thick green circle in Fig.~\ref{fig:clsd:image}) as an
overdensity of galaxies with a similar colour of R$-$z'$=1.4^m$,
and cluster optical luminosity function compatible with $z=0.9$ (Andreon et al. \cite{and04}).

The peak in the redshift distribution at $z\approx 0.68$ (see
Fig.~\ref{fig:redshifts}) does not correspond to any spatial grouping.

\underline{\em \clsE\/:} This is a very interesting and complicated case. The
morphology of the X-ray emission suggests a bi-modality, with a
central extended source and a possible subclump to the south-east (see
Fig.~\ref{fig:clse:image}) . It is not trivial, however, to
disentangle the centre, where galaxies at $z=0.92-0.96$ and
$z=1.02-1.06$ are mixed.  The redshift space distance is rather
significant: $\Delta z \sim 0.1$ ($\sim 15000$ km s$^{-1}$) which
rules out the galaxies belonging to one and the same cluster.

This cluster has been observed in the K' band with ESO-NTT/SOFI in the
framework of the XMM-LSS VIRMOS Deep Survey over an area of 0.25
deg$^2$ (Iovino et al. 2004, in preparation).  The exposure of
1.5 hours resulted in a catalog down to a limiting magnitude
K'$_{AB}\approx22.7$. A search for overdensities in photometric
redshift slices of $\Delta z_{phot}=0.2$ was made, with $z_{phot}$
calculated from BVRIK' photometry. As a result, there is a strong
detection with signal-to-noise (S/N) $\sim5$ of an overdensity at
$z\approx 0.9$ centred on the south-east extension of the X-ray
emission.  Observations in K at ESO-NTT/SOFI and R and z' at CTIO
(Andreon et al. \cite{and04}) also indicate two overdensities of
galaxies with colours compatible with $z\approx 1$.

The large fraction of emission line objects is also striking. From 5
objects near the X-ray peak with redshifts between $0.92-1.06$, 3 have
detectable [OII]3727\AA\ emission in their spectra. As in the case of
\clsC, it is difficult to assess the presence of an AGN from this one
emission line. At these redshifts, broad [MgII]2800\AA\ would be
detectable, but in no case was it found.

Based on the X-ray-to-optical flux ratios it is possible that some
contribution to the overall X-ray emission might be due to AGN, but
unlikely that all of it is due to AGN.

Our preliminary interpretation is that we are observing two clusters
in projection: one at $z\sim 0.93$, forming the main body of the
extended X-ray source, and the more distant one at $z=1.02$, that is
partially covered.  The redshifts of the spectroscopically observed
objects inside the X-ray emission span all of the interval from $z
\sim 0.92$ to $z\sim 1.05$. This could be an indication of that some
of these galaxies may lie on a filament connecting both
clusters. Their number, however, is insufficient for any strong claim.

Note that \clsE\ is very similar to the case of RXJ1053.7+5735
(Hasinger et al. \cite{has98}) -- a double cluster at $z=1.14$
(Hashimoto et al. \cite{has04}), which was repeatedly reported in the
literature at $z=1.26$, mistakenly taken a cD-like galaxy from the
background as the cluster central galaxy. Thanks to the deepest XMM
single pointing observation it was possible to disentangle
RXJ1053.7+5735. Unfortunately we cannot apply the same technique as
the exposure time in our case is $\sim 50$ times shorter. \clsE\ is
however a good target for deep integral field spectroscopy
observations (e.g.IFU of the VIMOS instrument) that will help us
resolve this extremely interesting case.

\begin{figure}[ht]
  \caption[]{\clsA. CFH12k I-band image of a $7\arcmin \times
  7\arcmin$ field is shown. North is up and east to the left.  The
  X-ray contours run from 0.1 to 5 counts/pixel ($2.5\arcsec$ pixel
  size) with 10 levels in log space. Objects in the $0.6 \leq z \leq
  0.63$ redshift range are denoted by boxes and triangles (for
  emission-line objects). } \label{fig:clsa:image}
\end{figure}

\begin{figure}[ht]
	\caption[]{\clsB. A
  $7\arcmin \times 7\arcmin$ field is shown and the objects at $0.76
  \leq z \leq 0.81$ are denoted by boxes and triangles (for
  emission-line objects). The X-ray contours run from 0.1 to 5
  photons/pixel with 10 levels in log space, with the lowest cluster
  contour at 0.1 photons/pixel. } \label{fig:clsb:image}
\end{figure}

\begin{figure}[ht]
\caption[]{\clsC. Only
  the central $5\arcmin \times 5\arcmin$ region of the field is shown
  and the objects in $0.82 \leq z \leq 0.86$ are denoted by boxes and
  triangles (for emission-line objects). The X-ray contours run from
  0.1 to 5 photons/pixel with 10 levels in log space, and the lowest
  cluster contour is at 0.1 photons/pixel.}  \label{fig:clsc:image}
\end{figure}

\begin{figure}[ht]
\caption[]{\clsD. Only the
  central $3.4\arcmin \times 3.4\arcmin$ region of the field is shown
  and the redshifts for the spectroscopically observed objects at
  $z>0.8$ are indicated. The triangles denote emission-line
  objects. The X-ray contours run from 0.15 to 5 photons/pixel with 10
  levels in log space, with the lowest cluster contour at 0.15
  photons/pixel. The optical detection from CTIO Rz' data is shown as
  a green circle.}  \label{fig:clsd:image}
\end{figure}

\begin{figure}[ht]
\caption[]{\clsE. Only the
  central $3\arcmin \times 3\arcmin$ region of the field is shown and
  the redshifts for the spectroscopically observed objects at $z>0.2$
  are indicated. The triangles denote emission-line objects. The X-ray
  contours run from 0.15 to 5 photons/pixel with 10 levels in log
  space, with the lowest cluster contour at 0.15 photons/pixel.}
  \label{fig:clse:image}
\end{figure}

\section{Conclusions}
\label{sec:end}

We present five newly discovered high redshift, X-ray selected
clusters in the XMM Large-Scale Structure Survey. The detection and
classification in X-rays, the subsequent optical identification and
spectroscopic target selection demonstrate the efficiency of the
programme and the associated follow-up.

\subsection{Highlights}

\begin{itemize}
\item The five newly discovered X-ray clusters at $z>0.6$ were observed at
  ESO-VLT for 11.5 hours in total. For three of them, we have measured
  more than 15 galaxies with concordant redshifts and thus obtained a
  viable estimate of the velocity dispersion.
\item Thanks to the optimised target selection, we were able to get an
  estimate of the velocity dispersion of \clsC\ at $z=0.84$ in a
  reasonable VLT time of 2 hours. To show the potential of the XMM-LSS
  it is interesting to note that to date there are 5 known clusters in
  the literature (two of them are X-ray selected) at $z>0.8$ with more
  than 10 spectroscopically measured concordant redshifts.
\item We have detected a complex structure at a redshift of unity --
  \clsE. The X-ray morphology and the presence of galaxy overdensity
  in photometric redshift space and in K, R and z' colour space allow
  us to speculate that most likely we see two clusters in
  projection. Some galaxies included in the spectroscopic observations
  may reside in a possible filament connecting the two clusters.
\item From CTIO Rz' observations, \clsD\  is detected as an overdensity of
  R$-$z'=$1.4^m$ galaxies, matching the redshift $z\sim 0.9$ of two
  bright and centrally located galaxies inside the X-ray extended
  emission.
\item From Tables~\ref{tab:x1} and \ref{tab:x2}, and
  Fig.~\ref{fig:scaling}, it is obvious that for the first time we are
  unveiling, in a systematic manner, moderate mass clusters in
  the redshift range $0.5<z<1$.  Most of these objects are weak,
  extended X-ray sources and consequently would have been difficult to
  classify in typical deep ROSAT/PSPC pointings because of the
  insufficient photon statistics in addition to the worse PSF,
  although there are successful ROSAT identifications of a handful of
  clusters at similar redshifts and mass scales (eg. Maughan et
  al. \cite{mau03}, Vikhlinin et al. \cite{vik02}). Thus, with
  XMM-LSS we are starting to fill the cluster database with 
  significant number of objects at high redshift from the middle of
  the mass function. This is a great improvement upon ROSAT based
  cluster surveys and will bring important new information on
  cosmological constraints as well as on non-gravitational physics in
  clusters (as the effects like pre-heating and feedback are
  presumably more important in lower mass systems).
\end{itemize}

\subsection{Prospects}

In the near future, our goal will be to confirm, via spectroscopic
identifications, all XMM-LSS clusters down to $\sim 8\times 10^{-15}$
\flux\ over $\sim 8$ sq.deg. of the AO-1 and AO-2 observations. This
will form a complete sample of about 100 X-ray selected clusters at
$0<z<1$.  A follow-up of the most interesting objects at $z\sim 1$ is
also previewed with integral-field spectroscopy and possibly deeper
targeted X-ray observations with XMM or Chandra for better insight
into the cluster physics.

\begin{acknowledgements}
 
  MP and IV are grateful to the ESO/Santiago Office for Science, for a
  two week stay in October 2002, where the analysis of the VLT data
  presented here was initiated. We are very thankful for the help by
  G. Marconi and the ESO/VLT team during the observing run. SDS is
  supported by a post-doctoral position from the Centre National
  d'Etudes Spatiales.  We wish to thank the referee, H. Ebeling, for
  his critical approach that helped to improve the paper.

  This research has made use of the X-Rays Clusters Database (BAX)
  which is operated by the Laboratoire d'Astrophysique de
  Midi-Pyr\'en\'ees (LAOMP), under contract with the Centre National
  d'Etudes Spatiales (CNES) and of the NASA/IPAC Extragalactic
  Database (NED) which is operated by the Jet Propulsion Laboratory,
  California Institute of Technology, under contract with the National
  Aeronautics and Space Administration.

\end{acknowledgements}

\end{document}